\documentclass{article}
\usepackage{spconf,amsmath,graphicx}

\usepackage{xcolor}
\usepackage[hidelinks]{hyperref}
\usepackage{mathabx}
\usepackage{flushend}

\usepackage{fancyhdr}
\fancypagestyle{firstpage}{% Page style for first page
  \fancyhf{}% Clear header/footer
  \setlength{\headheight}{15mm}
  \renewcommand{\topmargin}{-15mm}
  \fancyhead[C]{\vspace{5mm} \small \textit{IEEE ICASSP 2023 Workshop on Humans, Machines and Multimedia - Quality of Experience and Beyond}}
}

\newcommand{\etal}{\textit{et al.}}

\title{LCCM-VC: Learned Conditional Coding Modes for Video Compression
}
\name{Hadi Hadizadeh$^{1,2}$ and Ivan V. Baji\'{c}$^{\;2}$\thanks{This work was supported in part by NSERC grants RGPIN-2021-02485 and RGPAS-2021-00038. %H. Hadizadeh is also with Quchan University of Technology (email: h.hadizadeh@qiet.ac.ir). 
}
}
\address{$^{1}$ Quchan University of Technology, Iran \hspace{15pt} $^{2}$ Simon Fraser University, Canada}
\begin{document}
%\ninept
%
\maketitle
\begin{abstract}
End-to-end learning-based video compression has made steady progress over the last several years. However, unlike learning-based image coding, which has already surpassed its handcrafted counterparts, the progress on learning-based video coding has been slower. In this paper, we present learned conditional coding modes for video coding (LCCM-VC), a video coding model that competes favorably against HEVC reference software implementation. Our model utilizes conditional -- rather than residual -- coding, and introduces additional coding modes to improve compression performance. The compression efficiency is especially good in the high-quality/high-bitrate range, which is important for broadcast and video-on-demand streaming applications. The implementation of LCCM-VC is available at \url{https://github.com/hadihdz/lccm_vc}
\end{abstract}
\begin{keywords}
End-to-end learned video coding, conditional coding, augmented normalizing flows, autoencoders
\end{keywords}

\thispagestyle{firstpage}

\section{Introduction}
\label{sec:intro}

With the rapid advancement of deep learning technologies, 
various end-to-end learned image/video codecs have been developed~\cite{balle2018, minnen, cheng2020} to rival their handcrafted %non-neural
counterparts such as JPEG, High Efficiency Video Coding (HEVC)~\cite{hevc}, and Versatile Video Coding (VVC)~\cite{VVC}. For instance, in the seminal work by Ball\'{e} \etal~\cite{balle2018}, the variational autoencoders (VAE) were used to construct an end-to-end learned image compression system based on a context-adaptive entropy model. This model incorporates a hyperprior as side information to effectively capture dependencies in the latent representation, thereby improving entropy modeling. Many follow-up VAE-based models were then developed to further improve compression performance~\cite{minnen, lee, cheng2020}. A popular one 
by Cheng \etal~\cite{cheng2020} used discretized Gaussian mixture likelihoods to parameterize the latent distribution 
for  
entropy modeling, achieving high rate-distortion (RD) performance.
In fact, the results in~\cite{cheng2020} show that this model achieves superior performance on both PSNR and MS-SSIM quality metrics over 
JPEG, JPEG2000, and HEVC (Intra), and comparable performance with 
VVC (Intra). 

Although VAEs have been proven to be effective for image compression, 
their ability to provide a wide range of bitrates and  reconstruction qualities has been called into question~\cite{ae_limit}. 
To address this issue, an image compression method was proposed in~\cite{ae_limit} based on normalizing flows. Using augmented normalizing flows (ANF), Ho \etal~\cite{anfic} developed ANF for image compression (ANFIC), 
which combines both VAEs and normalizing flows to achieve the state-of-the-art performance for image compression, even better than \cite{cheng2020}.

%\vspace{-1pt}
Building on the success of learned image compression, learned video compression is catching up quickly. Lu \etal~\cite{dvc} presented deep video compression (DVC)  
as the first end-to-end learned video codec based on temporal predictive coding. Agustsson \etal~\cite{ssf} proposed an end-to-end video coding model based on a learning-based motion compensation framework in which a warped frame produced by a learned flow map is used a predictor for coding the current video frame. Liu \etal~\cite{liu} used feature-domain warping in a coarse-to-fine manner for video compression. Hu \etal~\cite{hu} employed deformable convolutions for feature warping.

\begin{figure*}
\centering
\includegraphics[scale=0.4]{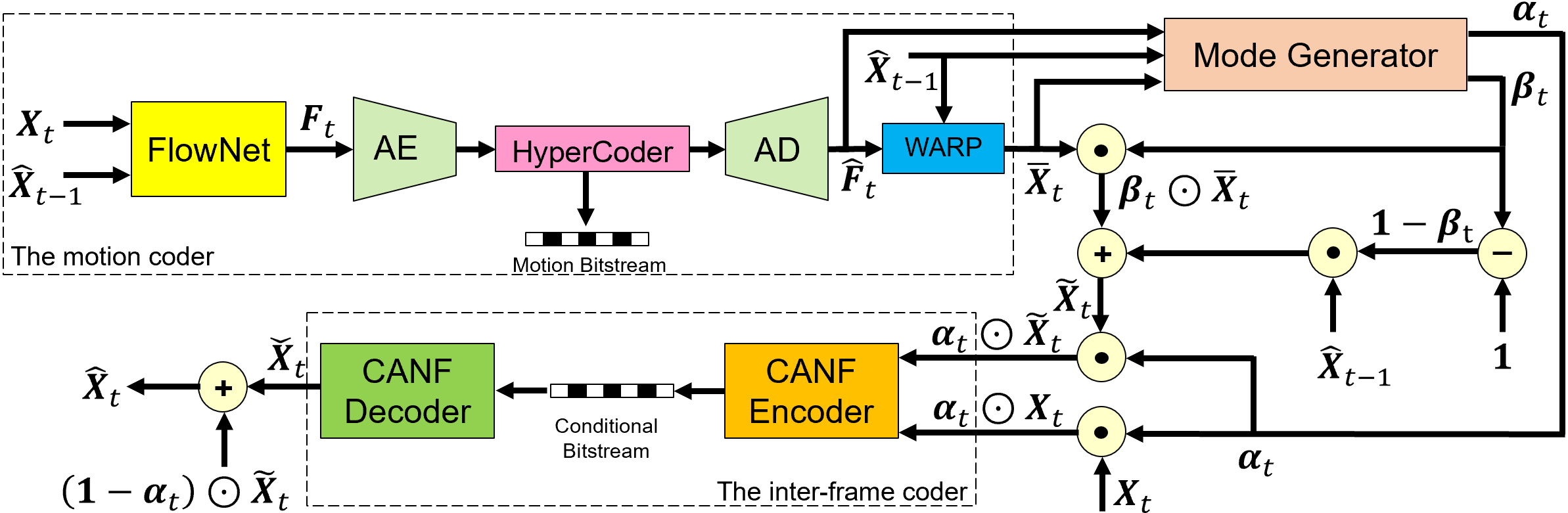}
\caption{The block diagram of the proposed learned video compression system. AE and AD are the encoder and decoder of the HyperPrior coder from~\cite{balle2018}, respectively, and $\odot$ is point-wise multiplication. What is shown is the encoding of the first P frame where we use the HyperPrior coder for motion coding. We use PWC-Net~\cite{pwc} as the FlowNet.}
\label{fig:flowchart1}
\end{figure*}

%\vspace{-1pt}
Most of the existing video codecs rely on residual coding. However, Ladune \etal~\cite{theo2,ladune} argued that conditional coding relative to a predictor is more efficient than residual coding using the same predictor. 
Building on this idea, Ho \etal~\cite{canf} proposed conditional augmented normalizing flows for video coding (CANF-VC), 
which achieves very strong performance among learned video codecs. CANF-VC uses conditional coding for both motion and inter-frame coding. 

%\vspace{-1pt}

Here we extend these ideas further. We provide a more comprehensive theoretical justification for conditional coding relative to multiple predictors/modes, and construct a codec called Learned Conditional Coding Modes for Video Compression (LCCM-VC), 
which outperforms representative learning-based codecs 
on the commonly used test sets.  
The proposed system is described in Section \ref{sec:proposed}. 
Experiments are presented in Section~\ref{sec:experiments}, followed by
conclusions 
in Section~\ref{sec:conclusions}.

\vspace{-10pt}
\section{Proposed Method}
\label{sec:proposed}
\vspace{-5pt}
\subsection{Motivation}
\label{sec:motivation}
\vspace{-3pt}
Let $X$ and $Y$ be two random variables and $R=X-Y$, then 
\begin{equation}
\begin{split}
    H(X|Y) = H(R+Y|Y) &\stackrel{\text{(a)}}{=} 
    H(R|Y) \\
    &\stackrel{\text{(b)}}{\leq} H(R) = H(X-Y),
\end{split}
\label{eq:conditional_vs_residual}
\end{equation}
where $H(\cdot)$ is the entropy, $H(\cdot|\cdot)$ is the conditional entropy, (a) follows from the fact that given $Y$, the only uncertainty in $R+Y$ is due to $R$, and (b) follows from the fact that conditioning does not increase entropy~\cite{Cover_Thomas_2006}.

Now consider the following Markov chain $X \to Y \to f(Y)$ where $f(\cdot)$ is an arbitrary function. By the data processing inequality~\cite{Cover_Thomas_2006}, we have $I(X;f(Y))\leq I(X;Y)$, where $I(\cdot;\cdot)$ is the mutual information. Expanding the two mutual informations as follows: $I(X;f(Y)) = H(X) - H(X|f(Y))$ and $I(X;Y)=H(X)-H(X|Y)$, and applying the data processing inequality, we conclude
\begin{equation}
    H(X|Y) \leq H(X|f(Y)).
    \label{eq:cond_entropy_f}
\end{equation}

In video compression, coding modes are constructed via predictors, for example inter coding modes use other frames to predict the current frame, while intra coding modes use information from the same frame for prediction. Let $X=\mathbf{X}_t$ be the current frame,  $Y=\{\mathbf{X}^{(1)}, ..., \mathbf{X}^{(n)}\}$ be a set of $n$ candidate predictors, and $\mathbf{X}_p=f(\mathbf{X}^{(1)}, ..., \mathbf{X}^{(n)})$ be a predictor for $\mathbf{X}_t$ from $\{\mathbf{X}^{(1)}, ..., \mathbf{X}^{(n)}\}$. Function $f(\cdot)$ could, for example, use  different combinations of $\{\mathbf{X}^{(1)}, ..., \mathbf{X}^{(n)}\}$ in different regions of the frame. Further, let $f^*(\mathbf{X}^{(1)}, ..., \mathbf{X}^{(n)})$ be an optimal predictor for $\mathbf{X}_t$ that minimizes conditional entropy. Then, based on~(\ref{eq:conditional_vs_residual}) and~(\ref{eq:cond_entropy_f}), 
\begin{equation}
\begin{split}
    H(\mathbf{X}_t|\mathbf{X}^{(1)}, ..., \mathbf{X}^{(n)}) &\stackrel{\text{(a)}}{\leq} H(\mathbf{X}_t|f^*(\mathbf{X}^{(1)}, ..., \mathbf{X}^{(n)})) \\
    &\stackrel{\text{(b)}}{\leq} H(\mathbf{X}_t|f(\mathbf{X}^{(1)}, ..., \mathbf{X}^{(n)})) \\
    &= H(\mathbf{X}_t|\mathbf{X}_p) \stackrel{\text{(c)}}{\leq} H(\mathbf{X}_t - \mathbf{X}_p),
\end{split}
\label{eq:cond_entropy_n_pred}
\end{equation}
where (a) follows from~(\ref{eq:cond_entropy_f}), (b) follows from the fact that $f^*(\cdot)$ is the optimal predictor, and (c) follows from~(\ref{eq:conditional_vs_residual}). Also note that if $m>n$, then 
\begin{equation}
    H(\mathbf{X}_t|f^*(\mathbf{X}^{(1)}, ..., \mathbf{X}^{(m)})) \leq H(\mathbf{X}_t|f^*(\mathbf{X}^{(1)}, ..., \mathbf{X}^{(n)})),
\label{eq:more_predictors}
\end{equation}
because an optimal predictor with more candidates ($m$) can, at the very least, choose to ignore $m-n$ of them and achieve the same performance as the predictor with $n$ candidates. 

Our proposed codec is built on the idea shown in~(\ref{eq:more_predictors}). By generating a large number of candidate predictors (modes), we want to minimize the bitrate needed for coding $\mathbf{X}_t$ conditioned on an optimal combination of these modes. Note that the theory promises even better performance via multi-conditional coding: inequality (a) in~(\ref{eq:cond_entropy_n_pred}). However, this requires estimating conditional probabilities in very high-dimensional spaces, and is not pursued in this paper.

\subsection{Codec description}
\label{sec:codec_description}
Fig.~\ref{fig:flowchart1} depicts the structure of our proposed video compression system. It consists of three main components: 1) motion coder, 2) mode generator, and 3) inter-frame coder. The exact functionality of each of these components is described below.

\textbf{Motion coder:} Given the current frame  $\mathbf{X}_{t}$ and its reconstructed reference frame $\widehat{\mathbf{X}}_{t-1}$, we first feed them to a learned optical flow estimation network like PWC-Net~\cite{pwc}, to obtain a motion flow map $\mathbf{F}_{t}$. The obtained flow is then encoded by the encoder (AE) of the HyperPrior-based coder from~\cite{balle2018}, and the obtained motion bitstream is transmitted to the decoder. At the decoder side, the transmitted flow map is reconstructed by the decoder (AD) of the HyperPrior coder to obtain $\widehat{\mathbf{F}}_{t}$. Then, $\widehat{\mathbf{X}}_{t-1}$ is warped by $\widehat{\mathbf{F}}_{t}$ using bilinear sampling \cite{canf} to obtain a motion-compensated frame $\overline{\mathbf{X}}_{t}$.

\begin{figure}[t!]
\centering
\includegraphics[scale=0.4]{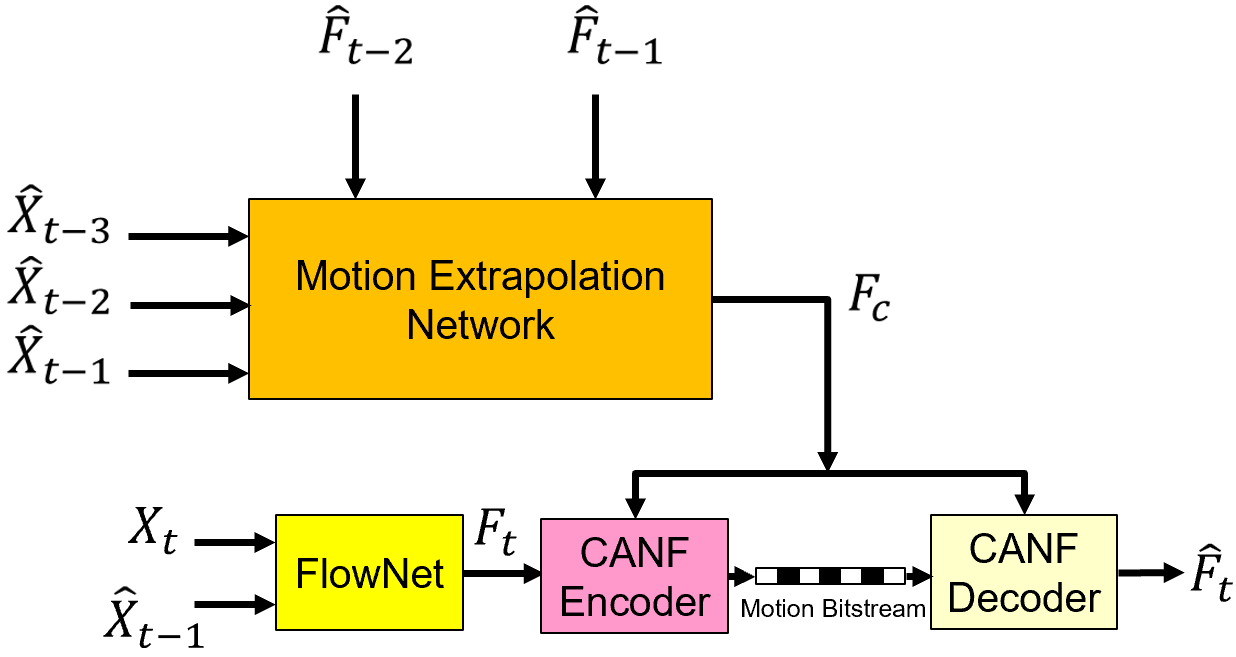}
\caption{The overall structure of the motion extrapolation network for producing a conditional flow, $\mathbf{F}_c$, for encoding $\mathbf{F}_t$.}
\label{fig:extrapolation}
\end{figure}

The above-described motion coder is only used for the first P frame in each  group of pictures (GOP). For the subsequent P frames, we use the CANF-based motion coder shown in Fig.~\ref{fig:extrapolation}. Here, the extrapolation network from CANF-VC~\cite{canf} is used to extrapolate a flow map $\mathbf{F}_c$ from the three previously-decoded frames $\widehat{\mathbf{X}}_{t-3}$, $\widehat{\mathbf{X}}_{t-2}$, $\widehat{\mathbf{X}}_{t-1}$ and two decoded-flow maps $\widehat{\mathbf{F}}_{t-2}$, $\widehat{\mathbf{F}}_{t-1}$. $\mathbf{F}_t$ is then coded conditionally relative to $\mathbf{F}_c$.  
I-frames are coded using ANFIC \cite{anfic}.

\textbf{Mode generator:} The goal of the mode generator is to produce additional coding modes to operationalize~(\ref{eq:more_predictors}).  
For this purpose, the previous reconstructed frame $\widehat{\mathbf{X}}_{t-1}$, the motion-compensated frame $\widehat{\mathbf{X}}_{t}$, and the decoded flow map $\widehat{\mathbf{F}}_t$ are concatenated and fed to the mode generator to produce two weight maps $\boldsymbol{\alpha}_t$ and $\boldsymbol{\beta}_t$, which are of the same size as the input image. The mode generator is implemented as a simple convolutional network with structure $[C_1,R_1,C_2]$, where $C_1$ and $C_2$ are  convolutional layers with 32 kernels of size $3\times3$ (stride=1, padding=1), and $R_1$ is a \texttt{LeakyReLU} layer whose negative slope is $0.1$. 
Since  
$\boldsymbol{\alpha}_t$ and $\boldsymbol{\beta}_t$ are produced from previously (de)coded data, they can be regenerated at the decoder without any additional bits. 
These two maps are then fed to a sigmoid layer to bound their values between 0 and 1. Then,  
a frame predictor $\widetilde{\mathbf{X}}_{t}$ is generated as:
\begin{equation}
\widetilde{\mathbf{X}}_{t} = \boldsymbol{\beta}_t \odot \overline{\mathbf{X}}_{t} +  (\mathbf{1}-\boldsymbol{\beta}_t) \odot \widehat{\mathbf{X}}_{t-1},
\end{equation}
where $\odot$ denotes Hadamard (element-wise) product and $\mathbf{1}$ is the all-ones matrix.
Moreover, $\boldsymbol{\alpha}_t$ is multiplied by both $\widetilde{\mathbf{X}}_{t}$ and $\mathbf{X}_{t}$, and the resultant two frames, $\boldsymbol{\alpha}_t \odot \widetilde{\mathbf{X}}_{t}$ and $\boldsymbol{\alpha}_t \odot \mathbf{X}_{t}$, are fed to the CANF-based conditional inter-frame coder for coding $\boldsymbol{\alpha}_t \odot \mathbf{X}_{t}$ conditioned on $\boldsymbol{\alpha}_t \odot \widetilde{\mathbf{X}}_{t}$. Note that $\boldsymbol{\alpha}_t$, $\boldsymbol{\beta}_t$, $\widehat{\mathbf{X}}_{t-1}$ and $\widetilde{\mathbf{X}}_{t}$ are available at the decoder.

%\subsection{The inter-frame coder}
\textbf{Inter-frame coder:} The inter-frame coder codes $\boldsymbol{\alpha}_t \odot \mathbf{X}_{t}$ conditioned on $\boldsymbol{\alpha}_t \odot \widetilde{\mathbf{X}}_{t}$ using the inter-frame coder of CANF-VC to obtain $\widecheck{\mathbf{X}}_t$ at the decoder. The final reconstruction of the current frame, i.e. $\widehat{\mathbf{X}}_{t}$, is then obtained by:
\begin{equation}
\widehat{\mathbf{X}}_{t} = \widecheck{\mathbf{X}}_t + (\mathbf{1}-\boldsymbol{\alpha}_t) \odot \widetilde{\mathbf{X}}_{t}.
\end{equation}

In the limiting case when $\boldsymbol{\beta}_t \to \mathbf{0}$, the predictor $\widetilde{\mathbf{X}}_{t}$  becomes equal to $\widehat{\mathbf{X}}_{t-1}$, and when $\boldsymbol{\beta}_t \to \mathbf{1}$, $\widetilde{\mathbf{X}}_{t}$  becomes equal to the motion-compensated frame $\overline{\mathbf{X}}_{t}$. For $\mathbf{0}<\boldsymbol{\beta}_t<\mathbf{1}$, the predictor $\widetilde{\mathbf{X}}_{t}$ is a pixel-wise mixture of $\widehat{\mathbf{X}}_{t-1}$ and $\overline{\mathbf{X}}_{t}$. Hence, $\boldsymbol{\beta}_t$ provides the system with more flexibility for choosing the predictor for each pixel within the current frame being coded. Also, for pixels where $\boldsymbol{\alpha}_t \to 0$, $\widehat{\mathbf{X}}_{t}$ becomes equal to $\widetilde{\mathbf{X}}_{t}$, so the inter-frame coder does not need to code anything. This resembles the SKIP mode in conventional coders, and depending on the value of $\boldsymbol{\beta}_t$, the system can directly copy from $\widehat{\mathbf{X}}_{t-1}$, $\overline{\mathbf{X}}_{t}$, or a mixture of these two, to obtain $\widehat{\mathbf{X}}_{t}$. When $\boldsymbol{\alpha}_t \to \mathbf{1}$, only the inter-frame coder is used to obtain $\widehat{\mathbf{X}}_{t}$. 
In the limiting case when $\boldsymbol{\alpha}_t \to \mathbf{1}$ and $\boldsymbol{\beta}_t \to \mathbf{1}$, the proposed method would reduce to CANF-VC~\cite{canf}. 

Note that a somewhat similar approach was proposed in~\cite{theo2}. However,~\cite{theo2} used only one weight map, which is similar to our~$\boldsymbol{\alpha}$ map, and this map was coded and transmitted to the decoder. 
In our proposed system, two maps,  $\boldsymbol{\alpha}$ and  $\boldsymbol{\beta}$, are used to create a larger number of modes. Moreover,  these two maps can be constructed using previously (de)coded information, so they can be regenerated at the decoder without any additional bits to signal  
the coding modes.

\begin{figure*}
\centering
\includegraphics[scale=0.28]{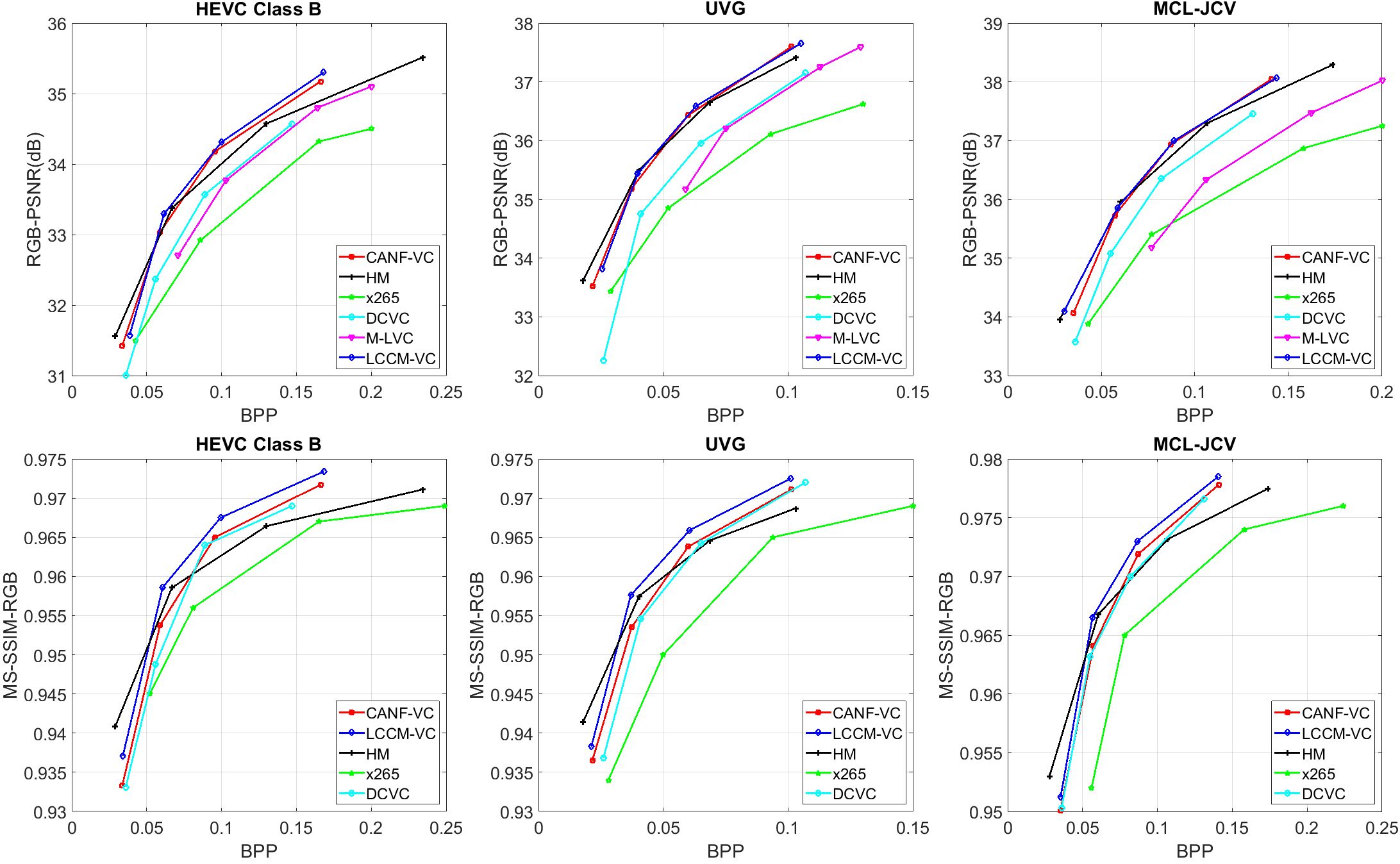}
\caption{Comparing various methods on three datasets: HEVC Class B, UVG, and MCL-JCV.}
\label{fig:results}
\vspace{-5pt}
\end{figure*}

\vspace{-5pt}
\section{Experiments}
\label{sec:experiments}
\vspace{-8pt}
\subsection{Training}
We trained the proposed LCCM-VC on the VIMEO-90K Setuplet dataset~\cite{vimeo}, which consists of 91,701 7-frame sequences with fixed resolution $448\times 256$, extracted from 39K  
video clips. We randomly cropped
these clips into $256\times 256$ patches, and used them for training LCCM-VC using a GOP of $N=5$ frames. We employed the Adam~\cite{adam} optimizer with the batch size of 4. We adopted a two-stage training scheme. In the first stage, we froze the CANF-based conditional coders with their pre-trained weights, and optimized the remainder of the model for 5 epochs with the initial learning rate of $10^{-4}$. In the second stage, we trained the entire system end-to-end for 5 more epochs with the initial learning rate of $10^{-5}$. Four separate models were trained for four different bitrates using the following  
loss function:
\begin{equation}
    \mathcal{L}= \sum_{i=1}^N \frac{\eta_i}{\sum_{j=1}^N \eta_j} \cdot \mathcal{L}_i,
\end{equation}
where $\eta_i = i$, and $\mathcal{L}_i$ is the RD loss of the $i$-th training frame defined in~\cite{canf} with $\lambda_1 \in \{256, 512, 1024, 2048\}$ and $\lambda_2 = 0.01 \cdot \lambda_1$. Note that~\cite{canf} used $\mathcal{L}=\sum_i \mathcal{L}_i$ as the training loss, without weighting. In our experiments, we first trained the model with $\lambda_1=2048$ (highest rate), and all  lower-rate models were then initialized from this model.
\vspace{-5pt}
\subsection{Evaluation methodology}
We evaluate the performance of LCCM-VC on three datasets commonly used in learning-based video coding: UVG \cite{uvg} (7 sequences), MCL-JCV \cite{mcl} (30 sequences), and HEVC Class B \cite{classb} (5 sequences). Following the common test protocol used in the recent literature~\cite{canf}, we encoded only the first 96 frames of the test videos, with the GOP size of 32. 

We used the following video codecs as benchmarks: x265 (`very slow' mode)~\cite{x265}, HEVC Test Model (HM 16.22) with LDP profile~\cite{hm} , M-LVC~\cite{mlvc}, DCVC~\cite{dcvc}, and CANF-VC~\cite{canf}.  
As the quality of the I-frames has a significant role on the RD performance of video codecs, in order to have a fair comparison, we used ANFIC~\cite{anfic} as the I-frame coder for all learned codecs in the experiment. Note that ANFIC achieves state-of-the-art performance for static image coding~\cite{anfic}. 

Similar to the existing practice in the learned video coding literature~\cite{canf, mlvc, dcvc}, to evaluate the RD performance of various methods, the bitrates were measured in bits per pixel (BPP) and the reconstruction quality was measured by both RGB-PSNR and RGB-MS-SSIM. Then the RD performance is summarized into BD-Rate~\cite{Bjontegaard}.

\vspace{-7pt}
\subsection{Results}
\label{sec:results}
In Fig.~\ref{fig:results}, we plot RGB-PSNR vs. BPP (top) and RGB-MS-SSIM vs. BPP curves (bottom) of various codecs on the three datasets. It is notable that both LCCM-VC and CANF-VC achieve better performance than HM 16.22 at higher bitrates/qualities, whereas HM 16.22 has slight advantage at lower bitrates. All three codecs 
offer comparable performance at medium bitrates. Tables~\ref{tab:bd_psnr} and~\ref{tab:bd_ssim} show BD-rate (\%) relative to the x265 anchor using RGB-PSNR and RGB-MS-SSIM, respectively, with negative values showing an average bit reduction (i.e., coding gain) relative to the anchor. The best result in each row 
is shown in blue, and the second best result is shown in red. As seen in the tables, LCCM-VC has the best RGB-PSNR results among learned codecs, and best RGB-MS-SSIM results overall.  

It should be mentioned that choosing x265 as the anchor -- in keeping with the established practice in learned video coding -- has a consequence of excluding high-quality portions of the RD curves from BD-Rate computation. Yet, high quality is where LCCM-VC (and CANF-VC) have the biggest advantage over HM 16.22, as seen in Fig.~\ref{fig:results}. In fact, if we choose HM 16.22 as the anchor, then LCCM-VC has BD-Rates of --2.8\%, 3.4\%, and --1.1\% (using RGB-PSNR), so in fact, it beats HM 16.22 on two out of three datasets even in terms of RGB-PSNR, and the gains are in the high-quality region.

\begin{table}[t]
\centering
\caption{BD-Rate (\%) relative to x265 using RGB-PSNR.}
\vspace{7pt}

\footnotesize
\setlength{\tabcolsep}{5pt}
\renewcommand{\arraystretch}{1.2}
\begin{tabular}{|c||c|c|c|c|c|}
    \hline
    Dataset & HM 16.22 & DCVC & M-LVC & CANF-VC & LCCM-VC \\
    \hline
    \hline
    HEVC-B & \textcolor{blue}{--32.1} & --10.5 & --9.7 & --27.7 & \textcolor{red}{--31.7} \\
    \hline
    UVG & \textcolor{blue}{--41.6} & --16.3 & --12.1 & --35.9 & \textcolor{red}{--36.6}\\
    \hline
    MCL-JCV & \textcolor{blue}{--38.6} & --21.3 & --5.3 & --32.0 & \textcolor{red}{--35.6} \\
    \hline
\end{tabular}
\label{tab:bd_psnr}
%\vspace{-5pt}
\end{table}

\begin{table}[t]
\centering
\caption{BD-Rate (\%) relative to x265 using RGB-MS-SSIM.}
\vspace{7pt}

\footnotesize
\setlength{\tabcolsep}{5pt}
\renewcommand{\arraystretch}{1.2}
\begin{tabular}{|c|c|c|c|c|}
    \hline
    Dataset & HM 16.22 & DCVC & CANF-VC & LCCM-VC \\
    \hline
    \hline
    HEVC-B & \textcolor{red}{--31.0}& --27.0 & --27.7 & \textcolor{blue}{--39.0} \\
    \hline
    UVG & --34.3 & --33.2&\textcolor{red}{--35.0} & \textcolor{blue}{--42.7}\\
    \hline
    MCL-JCV & --32.0& -36.1& \textcolor{red}{--36.9} & \textcolor{blue}{--49.7} \\
    \hline
\end{tabular}
\label{tab:bd_ssim}
%\vspace{-5pt}
\end{table}

\vspace{-5pt}
\section{Conclusions}
\label{sec:conclusions}
\vspace{-5pt}
In this paper, we proposed learned conditional coding modes for video coding (LCCM-VC), an  end-to-end learned video codec that  
achieves  
excellent results among learning-based codecs. We also gave a theoretical justification for advantages of conditional coding relative to multiple coding modes.  
LCCM-VC outperforms other benchmark codecs on three commonly used test datasets in terms of MS-SSIM and is competitive with HM 16.22 even in terms of RGB-PSNR.

\small
\bibliographystyle{IEEEbib}
\bibliography{strings,refs}

\begin{thebibliography}{10}

\bibitem{balle2018}
J.~Ball\'{e}, D.~Minnen, S.~Singh, S.~J. Hwang, and N.~Johnston,
\newblock ``Variational image compression with a scale hyperprior,''
\newblock in {\em Intl. Conf. on Learning Representations (ICLR)}, 2018, pp.
  1--23.

\bibitem{minnen}
D.~Minnen, J.~Ball\'{e}, and G.~D. Toderici,
\newblock ``Joint autoregressive and hierarchical priors for learned image
  compression,''
\newblock in {\em Advances in Neural Information Processing Systems}, 2018,
  vol.~31.

\bibitem{cheng2020}
Z.~Cheng, H.~Sun, M.~Takeuchi, and J.~Katto,
\newblock ``Learned image compression with discretized gaussian mixture
  likelihoods and attention modules,''
\newblock in {\em Proceedings of the IEEE Conference on Computer Vision and
  Pattern Recognition (CVPR)}, 2020.

\bibitem{hevc}
G.~J. Sullivan, J.~R. Ohm, W.~J. Han, and T.~Wiegand,
\newblock ``Overview of the high efficiency video coding {(HEVC)} standard,''
\newblock {\em IEEE Trans. Circuits and Systems for Video Technology}, vol. 22,
  no. 12, 2012.

\bibitem{VVC}
B.~Bross, Y.~K. Wang, Y.~Ye, S.~Liu, J.~Chen, G.~J. Sullivan, and J.~R. Ohm,
\newblock ``Overview of the versatile video coding ({VVC}) standard and its
  applications,''
\newblock {\em IEEE Trans. Circuits and Systems for Video Technology}, vol. 31,
  no. 10, 2021.

\bibitem{lee}
J.~Lee, S.~Cho, and S.~K. Beack,
\newblock ``Context-adaptive entropy model for end-to-end optimized image
  compression,''
\newblock in {\em Intl. Conf. on Learning Representations (ICLR)}, 2019.

\bibitem{ae_limit}
L.~Helminger, A.~Djelouah, M.~Gross, and C.~Schroers,
\newblock ``Lossy image compression with normalizing flows,''
\newblock in {\em Intl. Conf. on Learning Representations (ICLR)}, 2021.

\bibitem{anfic}
Y.~H. Ho, C.~C. Chan, W.~H. Peng, H~.M. Hang, and M.~Domanski,
\newblock ``{ANFIC}: Image compression using augmented normalizing flows,''
\newblock {\em IEEE Open Journal of Circuits and Systems}, vol. 2, pp.
  613--626, 2021.

\bibitem{dvc}
G.~Lu, W.~Ouyang, D.~Xu, X.~Zhang, C.~Cai, and Z.~Gao,
\newblock ``{DVC}: An end-to-end deep video compression framework,''
\newblock in {\em Proceedings of the IEEE/CVF Conference on Computer Vision and
  Pattern Recognition}, 2019, p. 11006–11015.

\bibitem{ssf}
E.~Agustsson, D.~Minnen, N.~Johnston, Ball\'{e} J., Hwang~S. J., and
  G.~Toderici,
\newblock ``Scale-space flow for end-to-end optimized video compression,''
\newblock in {\em Proceedings of the IEEE/CVF Conference on Computer Vision and
  Pattern Recognition}. IEEE, 2020, pp. 8503--8512.

\bibitem{liu}
H.~Liu, M.~Lu, Z.~Ma, F.~Wang, Z.~Xie, X.~Cao, and Y.~Wang,
\newblock ``Neural video coding using multiscale motion compensation and
  spatiotemporal context model,''
\newblock {\em IEEE Trans. Circuits and Systems for Video Technology}, 2020.

\bibitem{hu}
Z.~Hu, G.~Lu, and D.~Xu,
\newblock ``{FVC}: A new framework towards deep video compression in feature
  space,''
\newblock in {\em Proceedings of the IEEE/CVF Conference on Computer Vision and
  Pattern Recognition}, 2021, p. 1502–1511.

\bibitem{pwc}
D.~Sun, X.~Yang, M.Y. Liu, and J.~Kautz,
\newblock ``{PWC-N}et: {CNN}s for optical flow using pyramid, warping, and cost
  volume,''
\newblock in {\em Proceedings of the IEEE conference on computer vision and
  pattern recognition}, 2018, p. 8934–8943.

\bibitem{theo2}
T.~Ladune, P.~Philippe, W.~Hamidouche, L.~Zhang, and O.~Déforges,
\newblock ``Optical flow and mode selection for learning-based video coding,''
\newblock in {\em IEEE 22nd International Workshop on Multimedia Signal
  Processing,}, 2020.

\bibitem{ladune}
T.~Ladune, P.~Philippe, W.~Hamidouche, L.~Zhang, and O.~Deforges,
\newblock ``Conditional coding for flexible learned video compression,''
\newblock in {\em Neural Compression From Information Theory to
  Applications–Workshop@ ICLR 2021}, 2021.

\bibitem{canf}
Y.~H. Ho, C.~P. Chang, P.~Y. Chen, A.~Gnutti, and W.~H. Peng,
\newblock ``{CANF-VC}: Conditional augmented normalizing flows for video
  compression,''
\newblock {\em European Conference on Computer Vision}, 2022.

\bibitem{Cover_Thomas_2006}
T.~M. Cover and J.~A. Thomas,
\newblock {\em Elements of Information Theory},
\newblock Wiley, 2nd edition, 2006.

\bibitem{vimeo}
T.~Xue, B.~Chen, J.~Wu, D.~Wei, and W.T. Freeman,
\newblock ``Video enhancement with task-oriented flow,''
\newblock {\em International Journal of Computer Vision}, vol. 127, no. 8, pp.
  1106--1125, 2019.

\bibitem{adam}
D.~P. Kingma and J.~Ba,
\newblock ``Adam: A method for stochastic optimization,''
\newblock in {\em International Conference for Learning Representations}, 2015.

\bibitem{uvg}
A.~Mercat, M.~Viitanen, and J.~Vanne,
\newblock ``{UVG} dataset: 50/120fps 4k sequences for video codec analysis and
  development,''
\newblock in {\em Proceedings of the 11th ACM Multimedia Systems Conference},
  2020, pp. 297--302.

\bibitem{mcl}
H.~Wang, W.~Gan, S.~Hu, J.Y. Lin, L.~Jin, L.~Song, P.~Wang, I.~Katsavounidis,
  A.~Aaron, and C.C.J. Kuo,
\newblock ``{MCL-JCV}: a {JND}-based {H.264/AVC} video quality assessment
  dataset,''
\newblock in {\em 2016 IEEE International Conference on Image Processing
  (ICIP)}, 2016, pp. 1509--1513.

\bibitem{classb}
F.~Bossen,
\newblock ``Common test conditions and software reference configurations,''
\newblock in {\em JCTVC-L1100 12(7)}, 2013.

\bibitem{x265}
x265,
\newblock ``An open-source {HEVC} encoder,''
  \url{https://x265.readthedocs.io/en/master/}, 2022-03-10.

\bibitem{hm}
HM,
\newblock ``Reference software for {HEVC},''
  \url{https://vcgit.hhi.fraunhofer.de/Zhu/HM/blob/HM-16.22/cfg/encoder\_lowdelay\_P\_main.cfg},
  2022-03-10.

\bibitem{mlvc}
J.~Lin, D.~Liu, H.~Li, and F.~Wu,
\newblock ``{M-LVC}: multiple frames prediction for learned video
  compression,''
\newblock in {\em Proceedings of the IEEE/CVF Conference on Computer Vision and
  Pattern Recognition}, 2020, p. 3546–3554.

\bibitem{dcvc}
J.~Li, B.~Li, and Y.~Lu,
\newblock ``Deep contextual video compression,''
\newblock in {\em Advances in Neural Information Processing Systems}, 2021.

\bibitem{Bjontegaard}
G.~Bj\o{}ntegaard,
\newblock ``Calculation of average {PSNR} differences between {RD}-curves,''
  Apr. 2001,
\newblock VCEG-M33.

\end{thebibliography}

\end{document}